# Altermagnetic Multiferroics: Symmetry-Locked Magnetoelectric Coupling

Wei Sun[1], Changhong Yang[1], Xiaotian Wang[2], Shifeng Huang[1], Zhenxiang Cheng[2]*

[1]Shandong Provincial Key Laboratory of Green and Intelligent Building Materials, University of Jinan, Jinan, 250022, China

[2]Institute for Superconducting and Electronic Materials, Faculty of Engineering and Information Sciences, University of Wollongong, Squires Way, North Wollongong, NSW 2500, Australia

Multiferroics exhibit significant potential for low-power spintronic devices due to magnetoelectric coupling. Here, we discuss an emerging class of altermagnetic multiferroics, a system demonstrating distinct advantages including zero net magnetization (eliminating stray fields), momentum-dependent spin splitting (enabling electric-field control of spin currents), and intrinsic strong magnetoelectric coupling originating from spin-space symmetry.

**Overview of Multiferroics**

Multiferroic materials represent a unique class of functional materials characterized by the coexistence of multiple ferroic orders (e.g., ferroelectricity and ferromagnetism/ antiferromagnetism). Their technological significance stems from the intrinsic coupling between these order parameters, which enables electric-field control of magnetic properties-a crucial feature for developing next-generation, energy efficient multifunctional devices. Traditional classification schemes divide multiferroics into two fundamental categories: Type-I systems, where magnetic and ferroelectric orders originate independently, often exhibit robust individual ferroic properties but suffer from weak magnetoelectric coupling mediated through secondary mechanisms. Conversely, Type-II multiferroics derive their ferroelectricity directly from magnetic ordering, resulting in stronger magnetoelectric coupling, though typically at the expense of reduced polarization strength due to secondary nature of ferroelectricity in these systems.

At the microscopic level, spin-orbit coupling (SOC) serves as the primary mediator of magnetoelectric effects. SOC-dependent mechanisms like the Dzyaloshinskii-Moriya interaction facilitate real-space coupling between magnetic and ferroelectric order parameters. However, the inherent weakness of SOC (typically operating at meV energy scales) fundamentally limits the strength of magnetoelectric coupling and the resulting macroscopic polarization responses, creating a persistent bottleneck for practical applications.

The recent discovery of altermagnetism[1,2], provides a transformative solution to this longstanding challenge. This novel form of non-relativistic collinear magnetism creates a unique synergy between spin and spatial symmetries. Remarkably, while maintaining compensated magnetic order characteristic of antiferromagnets, altermagnets simultaneously exhibits spin-polarized behavior akin to ferromagnets, thereby reconciling properties previously considered mutually exclusive. This paradigm shift not only challenges conventional spin-only descriptions of magnetism but also introduces spatial symmetry as a critical design parameter, enabling direct coupling between spin configurations and ferroelectricity. This resulting spin space group formalism has given rise to a new class of materials with deterministic, intrinsic magnetoelectric multiferroicity.

Altermagnetic multiferroic offer several key advantages:

(1) The absence of net magnetic moment eliminates stray field interference common in ferromagnetic-based systems, while momentum-dependent spin splitting permits electric-field control of spin-polarized currents-ideal for high-density, low-power memory devices.

(2) The fundamental coupling between ferroelectricity and altermagnetic symmetry (described by spin space groups) creates a deterministic link between real-space electric polarization and momentum-space spin polarization, overcoming the weak coupling limitation of conventional Type-I and Type-II multiferroics.

(3) The magnetoelectric coupling originates not from weak SOC, but from the intrinsic structural interlocking inherent in the altermagnetic-ferroelectric order. This fundamental mechanism yields dramatically enhanced coupling strength and superior dynamic response characteristics, which are essential for practical spintronic applications.

The synergistic combination of these properties promises transformative advances in multiferroic materials and spintronic technologies. Unlike conventional approaches relying on real-space coupling between ferroelectric and magnetic orders, altermagnetic multiferroicity establishes an intrinsic momentum-space coupling between ferroelectric polarization and spin polarization. This fundamental distinction not only explains the superior performance characteristics but also reveals enormous potential for future information storage and processing technologies.

**Symmetry Conditions for Antiferromagnetism and Altermagnetism**

For collinear magnets, conventional antiferromagnets and altermagnets exhibit similar compensated magnetic orders in real space; their key distinction lies in momentum space. Within the framework of non-relativistic spin space groups, spin and spatial degrees of freedom are decoupled and can be represented as $[\mathcal{R}_i\|\mathcal{R}_j]$, where the operations on either side of the double vertical bar denote spin and spatial symmetry operations, respectively. If the system's Hamiltonian remains invariant under this symmetry, it transforms the energy eigenstates as $[\mathcal{R}_i\|\mathcal{R}_j]E(s, \boldsymbol{k}) = E(\mathcal{R}_i s, \mathcal{R}_j \boldsymbol{k})$. Furthermore, collinear magnets possess a $[\bar{\mathcal{C}}_2\|\mathcal{T}]$ symmetry. Here, $\bar{\mathcal{C}}_2$ represents the two-fold rotation-inversion operation, and $\mathcal{T}$ denotes the time-reversal operation. This symmetry enforces the relation $E(s, \boldsymbol{k}) = E(s, -\boldsymbol{k})$.

The spin degeneracy in conventional antiferromagnets stems from symmetries that relate sublattices through translation ($t$) or spatial inversion ($\mathcal{P}$) operations — though this connection is not explicitly accounted for in classical spin group theory. These symmetries, $[\mathcal{C}_2\|t]$ and $[\mathcal{C}_2\|\mathcal{P}]$, involve two-fold rotation ($\mathcal{C}_2$) around an axis perpendicular to the collinear spin axis (thereby flipping the spin s), while $\mathcal{P}$ and $t$ either reverse the wavevector $\boldsymbol{k}$ or leave it unchanged, respectively. The transformation properties of these symmetries are given by:

$$E(s, \boldsymbol{k}) = [\mathcal{C}_2\|t]E(s, \boldsymbol{k}) = E(-s, \boldsymbol{k}) \tag{1}$$

$$E(s, \boldsymbol{k}) = [\bar{\mathcal{C}}_2 || \mathcal{T}][\mathcal{C}_2 || \mathcal{P}]E(s, \boldsymbol{k}) = E(-s, \boldsymbol{k}) \tag{2}$$

Thus, the system exhibits complete spin degeneracy at every *k*-point.

In contrast, altermagnetism exhibits momentum-dependent spin polarization with an alternating distribution in *k*-space. Its emergence requires the system to preserve specific rotational or mirror symmetries that couple spin-opposite sublattices, denoted by the $[\mathcal{C}_2 || \mathcal{A}]$ symmetry. Here, $\mathcal{A}$ represents the set of symmetry operations satisfying altermagnetic constraints. Crucially, the system must break both $\mathcal{P}$ and $t$ symmetries to lift spin degeneracy. Notably, in 2D systems, $\mathcal{M}_z$ (mirror reflection along *z*-axis) and $\mathcal{C}_{2z}$ (two-fold rotation around *z*-axis) are functionally equivalent to $\mathcal{P}$ and $t$, respectively. Therefore, the symmetry conditions for 2D altermagnetism demand the simultaneous absence of $\mathcal{P}$, $t$, $\mathcal{M}_z$, and $\mathcal{C}_{2z}$ symmetries.

**Selective $\mathcal{P}$ or $t$ Symmetry-Breaking Driven Altermagnetic-Antiferromagnetic Phase Transitions**

When a system simultaneously satisfies the symmetry conditions for both antiferromagnetism and altermagnetism, the altermagnetic state is suppressed due to the energy eigenstates obeying $E(s, \boldsymbol{k}) = E(-s, \boldsymbol{k})$ for antiferromagnetism. Consequently, the system adopts an antiferromagnetic ground state with full spin degeneracy. To achieve magnetic order control, real-space $\mathcal{P}$ or $t$ symmetry breaking can induced via ferroelectric/antiferroelectric (FE/AFE) phase transitions. In FE/AFE systems, $\mathcal{P}$ symmetry is typically broken in the FE phase but restored in the AFE phase, whereas $t$ symmetry is usually broken in the AFE phase and restored in the FE phase. Crucially, since $\mathcal{P}$ is preserved in the AFE phase and $t$ is preserved in the FE phase, the key design principle for multiferroic materials requires breaking of either $\mathcal{P}$ or $t$ through an additional crystal field effect. This leads to two distinct material design strategies: (1) FE-driven $\mathcal{P}$-breaking: The FE phase breaks $\mathcal{P}$ symmetry, stabilizing an altermagnetic-ferroelectric state. (2) AFE-driven $t$-breaking: The AFM phase breaks $t$ symmetry, forming an altermagnet-antiferroelectric state (Fig. 1). Monolayer $CuCrP_2S_6$ (lacks $\mathcal{P}$ in both FE and AFE phases, $t$ is broken via FE to AFE transition) and stacked $SnS_2/MnPSe_3/SnS_2$ (lacks $t$ in both FE and AFE phases, $\mathcal{P}$ is broken via AFE to FE transition) are typical examples of this model.[3,4]

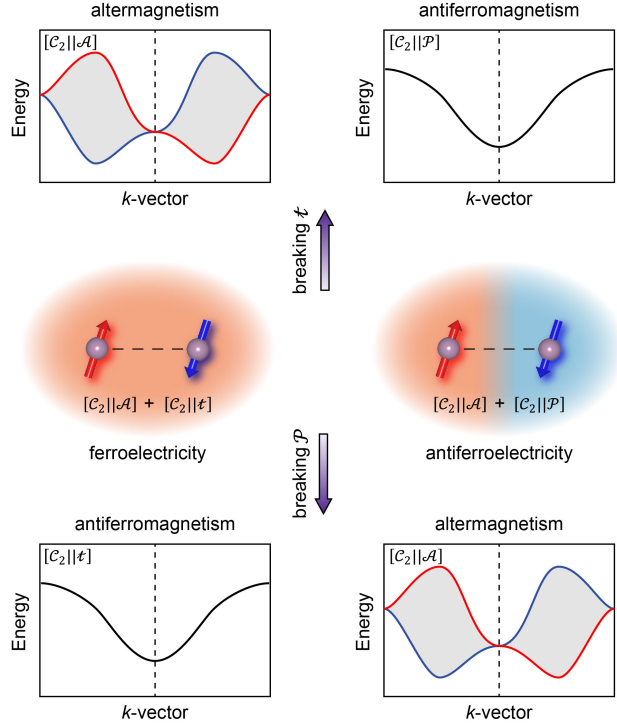

Fig. 1. Altermagnetic/antiferromagnetic phase transition induced by ferroelectric/antiferroelectric switching. The orange and cyan backgrounds represent opposite electric dipole moments, while the red and blue arrows denote opposite spins at magnetic sites, with the system exhibiting the $[\mathcal{C}_2\|\mathcal{A}]$ symmetry required for altermagnetism. Since ferroelectric and antiferroelectric orders typically exhibit $[\mathcal{C}_2\|t]$ or $[\mathcal{C}_2\|\mathcal{P}]$ symmetry, which suppresses altermagnetism, the breaking of either $t$ or $\mathcal{P}$ via an additional crystal field can induce altermagnetic and antiferromagnetic phase transitions.

This symmetry framework highlights the strong correlation between electric polarization and magnetic order: by choosing an FE/AFE phase transition to break specific spatial symmetries ($\mathcal{P}$ or $t$), the degeneracy of the magnetic ground state can be directionally controlled, offering a universal theoretical principle for multiferroic device design.

## $\mathcal{P}\mathcal{A}^{-1}$ Symmetry-Mediated Pseudo-Time-Reversal Effect

While the ferroelectric/antiferroelectric phase transitions offer unique advantages in strong magnetoelectric coupling and non-volatile control, the direct reversal of ferroelectric polarization stands out for its low energy consumption, ultrafast response, and high stability. This approach aligns particularly well with the core requirements of modern electronic devices. However, it imposes more stringent symmetrical conditions on the system.

Unlike the altermagnetic phase transitions described earlier, achieving this requires that altermagnetism not only coexist with ferroelectricity (rather than antiferroelectricity) but also that its spin-splitting direction be controlled by the polarization direction, as illustrated in Fig. 2. Gu, Liu et al., by screening the 2001 magnetic structures in the MAGNDATA database [21], identified 22 materials that satisfy the coexistence of altermagnetism and ferroelectricity. However, only 2 materials support the cooperative switching of ferroelectric polarization with the altermagnetic spin splitting.[5] Additionally, Smejkal and Zhou et al. proposed $BaCuF_4$ and $VOX_2$ as multiferroic materials achieving spin-ferroelectric interlock in 3D and 2D systems, respectively.[6,7]

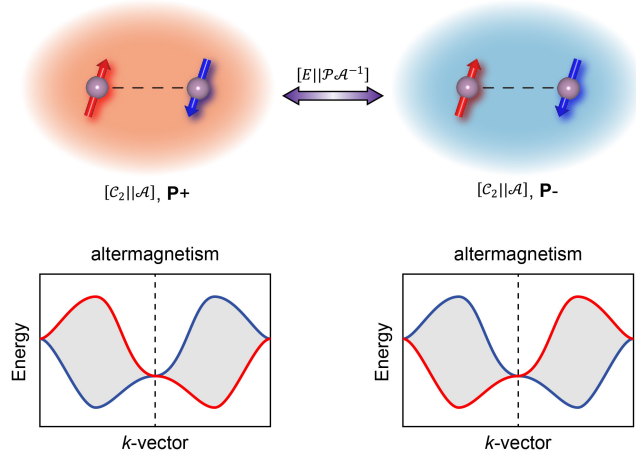

Fig. 2. Polarization-controlled reversal of altermagnetic spin splitting directions. The red and blue arrows denote opposite spins at magnetic sites under the $[\mathcal{C}_2\|\mathcal{A}]$ symmetry required for altermagnetism, while **P+** and **P-** represent opposite polarization states mappable via $[E\|\mathcal{P}\mathcal{A}^{-1}]$ symmetry. Under this symmetrical condition, polarization reversal directly controls the altermagnetic spin-splitting direction, which is equivalent to reversing the magnetic sites.

Sun, Cheng et al. proposed that when altermagnetism coexists with ferroelectricity under the symmetry $[\mathcal{C}_2\|\mathcal{A}]$, the condition for spin-ferroelectric interlock requires that ferroelectric switching be achieved via the combined spatial operation $\mathcal{P}\mathcal{A}^{-1}$ (denoted $[E\|\mathcal{P}\mathcal{A}^{-1}]$), rather than $\mathcal{P}$-only.[8] Here, $E$ represents the identity operation, $\mathcal{A}^{-1}$ denotes the inverse operation of the symmetry transformation $\mathcal{A}$, satisfying $\mathcal{A}\cdot\mathcal{A}^{-1}=E$. Under these conditions, the energy eigenstates transform into the altermagnetic $[\mathcal{C}_2\|\mathcal{A}]$ and ferroelectric phases $[E\|\mathcal{P}\mathcal{A}^{-1}]$ as follows:

$$[E\|\mathcal{P}\mathcal{A}^{-1}]E(s,k) = E(s,-\mathcal{A}^{-1}\boldsymbol{k}) \qquad (3)$$

$$[\mathcal{C}_2||\mathcal{A}]E(s,k) = E(-s, \mathcal{A}\boldsymbol{k}) \tag{4}$$

Thus, the combined symmetry $[E||\mathcal{P}\mathcal{A}^{-1}][\mathcal{C}_2||\mathcal{A}]$ yields:

$$[E||\mathcal{P}\mathcal{A}^{-1}][\mathcal{C}_2||\mathcal{A}]E(s,k) = E(-s,-\boldsymbol{k}) = \mathcal{T}E(s,k) \tag{5}$$

This indicates that ferroelectric reversal has the same effect on the energy eigenstates as the magnetic time-reversal operation $\mathcal{T}$. Consequently, the ferroelectric polarization direction can be tightly locked to the spin polarization direction via the intertwined real-space symmetry, effectively equivalent to a 180° magnetic reversal. Based on this principle, they further proposed a bilayer MnPSe$_3$ multiferroic. Due to its intrinsic, robust magnetoelectric coupling, this system constitutes a new type of multiferroic material (i.e., Type-III Multiferroics).

Furthermore, based on symmetry analysis, Liu et al. proposed that sliding ferroelectricity offers a promising avenue for designing such novel multiferroic materials.[9] They established a general symmetry rule for achieving sliding ferroelectric control of spin polarization in altermagnets, based on the crystallographic layer group of the constituent monolayer, the stacking operation, and the magnetic configurations of the bilayer.

**Spin-Symmetry-Governed Altermagnetic Type-II Multiferroics**

An alternative approach to achieving altermagnetic-ferroelectric coupling involves modifications to spin symmetry (magnetic ordering). Based on this mechanism, Yao's and Zhang's group independently developed frameworks for type-II multiferroics. These frameworks demonstrate that magnetic ordering can simultaneously break spin degeneracy to induce altermagnetism and generate ferroelectricity through exchange striction. Validation in materials like LiMnO$_2$ and MgFe$_2$N$_2$ supports this approach.[10,11] Crucially, this mechanism operates fundamentally differently from the previously described coupling scheme where ferroelectricity induces and controls altermagnetism via real-space symmetry. In this case, polarization serves as a secondary parameter governed by magnetic order, defining the characteristic behavior of Type-II multiferroicity.

**Discussion & Perspective**

The emergence of altermagnetic multiferroics marks a paradigm shift in magnetoelectric coupling research: transitioning from the traditional real-space coupling between ferroelectric polarization and magnetic ordering to the symmetry-governed manipulation of ferroelectric polarization and momentum-space spin polarization. While the field is still nascent, promising aspects coexist with significant challenges. The inherent flexibility of stacked 2D materials offers a versatile platform for engineering altermagnetic multiferroicity, where layer-selective symmetry breaking combined with sliding ferroelectricity provides a potent route to tailor magnetoelectric responses free from substrate constraints. Critically, the deterministic link between ferroelectricity and spin splitting inherent in this symmetry-driven approach enables unprecedented electric-field control of spin polarization, defining a novel class of multiferroics characterized by non-volatile, low-energy switching. However, key hurdles must be overcome. Fundamentally, the spin-splitting energy in these symmetry-broken altermagnetic multiferroics remains weaker than in intrinsic altermagnets, constraining potential spintronic efficiency. Furthermore, experimental progress faces bottlenecks, as realizing proposed structures demands atomic-precision synthesis and advanced in situ probes capable of resolving the critical momentum-space spin polarization. Compounding these challenges, achieving spin-splitting energies of $\approx 100$ meV or higher is typically required to effectively overcome room-temperature thermal fluctuations and ensure sufficiently stable spin-polarized states for practical device operation.